\documentclass[letterpaper,useAMS,usenatbib]{mn2e}
\usepackage{comment}
\usepackage{amssymb}   
\usepackage{amsmath}   
\usepackage{amssymb}  
\usepackage{bm}

\usepackage{xspace}
\usepackage{graphicx}
\usepackage{verbatim}

\usepackage[T1]{fontenc}
\usepackage{aecompl}

\newcommand{\be}{\begin{equation}}
\newcommand{\ee}{\end{equation}}

\usepackage[hyphens,spaces,obeyspaces]{url}
\urldef{\footurl}\url{https://github.com/STScI-MIRI/ImagingFluxCal/blob/main/model_fluxes.py}

\newcommand{\webbpsf}{{\tt webbPSF~}}

\title[JWST lensed quasar dark matter survey]{JWST lensed quasar dark matter survey I: Description and First Results.}

\author[Nierenberg et al.]{
A.~M.~Nierenberg$^{1}$ \thanks{\tt anierenberg@ucmerced.edu}, 
R.~E.~Keeley$^{1}$,
D.~Sluse$^{2}$,
D.~Gilman$^{3,4,5}$,
S.~Birrer$^{6}$,
T.~Treu$^{7}$,
\newauthor
K.~N.~Abazajian$^{8}$,
T.~Anguita$^{9,10}$,
A.~J.~Benson$^{11}$,
V.~N.~Bennert$^{12}$,
S.~G.~Djorgovski$^{13}$,
\newauthor
X.~ Du$^{7}$,
C.~D. Fassnacht$^{14}$,
S.~F.~Hoenig$^{15}$,
A.~Kusenko$^{7,16}$,
C.~Lemon$^{17}$,
M.~Malkan$^{7}$,
\newauthor 
V.~Motta$^{18}$,
L.~A.~Moustakas$^{19}$,
D.~Stern$^{19}$,
R.~H.~Wechsler$^{20,21,22}$
\medskip\\
$^1$  University of California, Merced, 5200 N Lake Road, Merced, CA 95341, USA \\
$^{2}$ STAR Institute, Quartier Agora - All\'e du six Ao\^ut, 19c B-4000 Li\`ege, Belgium \\
$^3$ Department of Astronomy $\&$ Astrophysics, University of Chicago, Chicago, IL 60637, USA \\
$^4$ Department of Astronomy and Astrophysics, University of Toronto, 50 St.\ George Street, Toronto, ON, M5S 3H4, Canada\\ 
$^5$ Brinson Fellow \\
$^6$ Department of Physics and Astronomy, Stony Brook University, Stony Brook, NY 11794, USA\\
$^7$ UCLA Physics \& Astronomy, 475 Portola Plaza, Los Angeles, CA 90095-1547, USA\\
$^{8}$ Department of Physics and Astronomy,  University of California, Irvine, CA 92697-4575, USA\\
$^{9}$ Instituto de Astrofisica, Departamento de Ciencias Fisicas, Universidad Andres Bello, Chile\\
$^{10}$ Millennium Institute of Astrophysics, Chile\\
$^{11}$ Carnegie Institution for Science, Pasadena CA 91101, USA \\
$^{12}$ Physics Department, California Polytechnic State University, San Luis Obispo, CA 93407, USA \\
$^{13}$ California Institute of Technology, Pasadena CA 91125, USA \\
$^{14}$ Department of Physics and Astronomy, UC Davis, 1 Shields Ave., Davis CA 95616 \\
$^{15}$ School of Physics and Astronomy, University of Southampton, Southampton SO17 1BJ, United Kingdom \\
$^{16}$ Kavli Institute for the Physics and Mathematics of the Universe (WPI), UTIAS, The University of Tokyo, Kashiwa, \\ Chiba 277-8583, Japan \\
$^{17}$Institute of Physics, Laboratory of Astrophysics, Ecole Polytechnique F\'ed\'erale de Lausanne (EPFL), Observatoire de Sauverny, 1290 Versoix, Switzerland \\
$^{18}$ Instituto de F\'{\i}sica y Astronom\'{\i}a, Universidad de Valpara\'{\i}so, Avda. Gran Breta\~na 1111, Valpara\'{\i}so, Chile \\
$^{19}$ Jet Propulsion Laboratory, California Institute of Technology, 4800 Oak Grove Dr, Pasadena, CA 91109\\
$^{20}$ Kavli Institute for Particle Astrophysics \& Cosmology, P.O. Box 2450, Stanford University, Stanford, CA 94305, USA \\
$^{21}$ Department of Physics, Stanford University, 382 Via Pueblo Mall, Stanford, CA 94305, USA \\
$^{22}$ SLAC National Accelerator Laboratory, Menlo Park, CA 94025, USA \\
}

\def\be{\begin{equation}}
\def\ee{\end{equation}}

\begin{document}
\pagerange{\pageref{firstpage}--\pageref{lastpage}}\pubyear{2023}

\maketitle           

\label{firstpage}

                      
\begin{abstract}
The flux ratios of gravitationally lensed quasars provide a powerful probe of the nature of dark matter. Importantly, these ratios are sensitive to small-scale structure, irrespective of the presence of baryons. This sensitivity may allow us to study the halo mass function even below the scales where galaxies form observable stars.
For accurate measurements, it is essential that the quasar's 
light is emitted from a physical region of the quasar with an angular scale of milli-arcseconds or larger; this minimizes microlensing effects by stars within the deflector. The warm dust region of quasars fits this criterion, as it has parsec-size physical scales and dominates the spectral energy distribution of quasars at wavelengths greater than 10$\mu$m. The JWST Mid-Infrared Instrument (MIRI) is adept at detecting redshifted light in this wavelength range, offering both the spatial resolution and sensitivity required for accurate gravitational lensing flux ratio measurements. Here, we introduce our survey designed to measure the warm dust flux ratios of 31 lensed quasars. We discuss the flux-ratio measurement technique and present results for the first target, DES J0405-3308. We find that we can measure the quasar warm dust flux ratios with 3\% precision. Our simulations suggest that this precision makes it feasible to detect the presence of 10$^7$ M$_\odot$ dark matter halos at cosmological distances. Such halos are expected to be completely dark in Cold Dark Matter models.

\end{abstract}

\setcounter{footnote}{1}

\section{Introduction}
\label{sec:intro}

Understanding the properties and behavior of dark matter (DM) is essential to our understanding of structure formation and galaxy formation. Its existence is currently our best model for 
the structure and evolution of the universe from scales ranging from the cosmic microwave background \citep{planck_collaboration_planck_2020} to the rotation curves of spiral galaxies and the dispersion support of spheroidal dwarf galaxies \citep[see, e.g.][and references therein]{weinberg_cold_2015, bullock_small-scale_2017}.  In this theory, baryonic galaxies form within extended dark matter halos \citep{white_core_1978, white_galaxy_1991}. Direct detection of these dark halos would provide robust evidence for dark matter's existence. Moreover, the particle properties of dark matter, such as its mass, formation mechanism, and possible self-interactions, determine the abundance and internal structure of halos \citep[see e.g.][and references therein]{buckley_gravitational_2018}. As dark matter continues to evade laboratory detection and is not guaranteed to be detected directly through non-gravitational interactions, observations of the properties of dark matter halos provide a crucial way
to test hypotheses about its particle properties. 

The `Cold' dark matter scenario and cosmological theory, $\Lambda \rm{CDM}$, predicts the existence of dark halos down to planet masses \citep{wang_universal_2020} in many models. Detecting these dark objects, below the expected scale of galaxy formation, would provide strong evidence in support of CDM and rule out entire classes of theories in which these low-mass objects do not exist. For example, warm dark matter (WDM) refers categorically to scenarios in which free-streaming suppresses the matter power spectrum below a characteristic scale, suppressing the concentration of halos and precluding their formation below a certain mass scale \citep{bode_halo_2001,schneider_non-linear_2012,bose_copernicus_2016,ludlow_mass-concentration-redshift_2016}. Self-interacting dark matter (SIDM) models introduce a self-interaction cross section between dark matter particles small enough to preserve the successes of CDM on large scales, but large enough to drive heat conduction through dark matter halos. This results in a dynamic evolution of halo density profiles that begins with core formation and eventual core collapse \citep{spergel_observational_2000,balberg_self-interacting_2002,kaplinghat_dark_2016,yang_strong_2023,yang_gravothermal_2023,zeng_core-collapse_2022}. Models in which an extremely light boson with a mass $\sim 10^{-22} \rm{eV}$ comprises all or part of the dark matter, usually referred to as "ultra-light dark matter" (ULDM) or fuzzy dark matter, predict suppression of small-scale structure similar to WDM, and manifest quantum-mechanical interference effects on galactic scales due to the kpc-scale de Broglie wavelength of the particles \citep{schive_cosmic_2014,mocz_galaxy_2017,chan_multiple_2020,laroche_quantum_2022,powell_lensed_2023}. More generally, any theory that modifies the linear matter power spectrum on scales $k > \rm{5 \ \rm{Mpc^{-1}}}$ impacts the abundance and internal structure of dark matter halos. This includes certain models of inflation, primordial non-Gaussianity, late-decaying DM particles, or a non-zero running spectral index in slow-roll inflation \citep{zentner_inflation_2002,stafford_bahamas_2020,gilman_primordial_2022,
ando_constraining_2022,maria_ezquiaga_massive_2022,esteban_milky_2023}. Primordial black holes (PBH) are another potential DM candidate that primarily affect the internal structure of subhalos \citep{afshordi_primordial_2003, ricotti_effect_2008, carr_primordial_2016,carr_primordial_2020,dike_strong_2023}

Galaxy-scale strong gravitational lensing can reveal dark matter structure through its gravitational effects on sub-galactic scales, and thus provide insight into its properties (see \citet{vegetti_strong_2023} for a comprehensive review). In a galaxy-scale strong gravitational lens, multiple images of a background source appear due to the deflection of light by a foreground galaxy and its surrounding dark matter halo. An extended background source, such as a galaxy, will appear warped and distorted by strong lensing, and will often partially encircle the foreground deflector. A more compact source, such as a quasar, typically appears two or four times from the perspective of the observer\footnote{If the source is a quasar surrounded by a galaxy, both extended arcs and multiple images of the quasar appear.}. The first derivative of the gravitational potential determines the relative positions of the lensed images, while the second derivative of the potential determines their magnifications. Thus, the positions and magnifications of lensed images constrain the mass distribution of the deflector across a range of scales, spanning the size of the Einstein radius (typically $\sim 1 \ \rm{arcsec}$) down to the milli-arcsecond scales probed by the image magnifications. These data are therefore sensitive to the abundance and internal structure of dark matter halos several orders of magnitude less massive than the main deflector and its host halo. The sensitivity of strong lensing observables to both the abundance and internal structure of halos has led to constraints on warm dark matter \citep{vegetti_constraining_2018,hsueh_sharp_2020,gilman_warm_2020,zelko_constraints_2022}, fuzzy dark matter \citep{laroche_quantum_2022,powell_lensed_2023}, self-interacting dark matter \citep{minor_unexpected_2021,gilman_strong_2021,gilman_constraining_2023}, primordial density fluctuations \citep{gilman_primordial_2022}, and primordial black holes \citep{dike_strong_2023}.

The state of the field has evolved considerably since \citet{mao_evidence_1998} and \citet{dalal_direct_2002} showed that low-mass dark matter halos could explain the relative magnifications (or flux ratios) of quadruply imaged radio-loud quasars. In the ensuing decades, the sample of known galaxy-scale strong lenses has grown by an order of magnitude, both through the discovery of new systems and the use of radio-quiet quasars observed at optical and infrared wavelengths. The modeling frameworks used to analyze and interpret data from strong lens systems now include more accurate models for the population of dark matter halos perturbing the lenses, including dark halos along the line of sight \citep{xu_effects_2012,despali_modelling_2018,gilman_probing_2019,gilman_constraints_2020,sengul_substructure_2022}, correlated structure around the host halo \citep{gilman_probing_2019}, and the tidal evolution of dark subhalos. The calibration of the substructure models implemented in lensing analyses come from the predictions of numerical simulations of structure formation in early-type galaxies \citep{fiacconi_cold_2016, nadler_symphony_2023} and semi-analytic models, including {\tt{galacticus}} \citep{benson_g_2012} and {\tt{SatGen}} \citep{jiang_satgen_2021}. Advances in the modeling of strong lens systems have been enabled by software packages such as {\tt{lensmodel}}\footnote{https://www.physics.rutgers.edu/\textasciitilde{}keeton/gravlens/}, {\tt{lenstronomy}}\footnote{https://github.com/lenstronomy/lenstronomy} \citep{birrer_lenstronomy_2018,birrer_lenstronomy_2021}, {\tt{GLEE}} \citep{suyu_halos_2010}, {\tt{PyAutoLens}}\footnote{https://github.com/Jammy2211/PyAutoLens}  \citep{nightingale_pyautolens_2021}, {\tt{Herculens}}\footnote{https://github.com/austinpeel/herculens} \citep{galan_herculens_2022}, and the codes of\citet{vegetti_bayesian_2009, vernardos_very_2022}, which include capabilities to forward-model lensing observables through multi-plane lensing computations and simultaneous reconstruction of lensed images and background sources. Finally, open-source packages such as {\tt{pyHalo}}\footnote{https://github.com/dangilman/pyHalo} and {\tt{paltas}}\footnote{https://github.com/swagnercarena/paltas}\citep{wagner-carena_images_2023} interface between lensing codes and dark matter models to quickly generate populations of dark matter halos for lensing simulations. 

The background source plays a key role in gravitational lensing inferences of dark matter structure from image flux ratios because its spatial extent imposes a particular angular and temporal scale on the problem. For substructure lensing studies, the source must be extended enough that the light-crossing time exceeds the arrival time difference between lensed images (typically days to months) so that intrinsic variations in the source produce a negligible change in the flux ratios. For a typical time delay of $\sim10$ days, this implies a spatial extent of at least 0.1 parsec. The source must also be extended enough to be insensitive to microlensing by stars in the main deflector. The perturbation of an image magnification caused by a halo depends on the deflection angle produced by the halo relative to the angular size of the source \citep{dobler_finite_2006,metcalf_small-scale_2012}. Stars produce deflection angles of order $\sim \mu$as. Given typical galaxy-scale lensing configurations, this implies a minimum required source size of $\sim$ mas, which corresponds to physical scales of $\sim$ 1 parsec at a typical source redshift of $z=2$. 
Quasar radio and narrow-line emission are extended enough to meet these criteria \citep{metcalf_compound_2001}, and these sources have yielded some of the strongest constraints to date on a turnover in the halo mass function  \citep{gilman_probing_2018,gilman_probing_2019,hsueh_sharp_2020}, with an upper limit of M$_{\rm{hm}} <10^{7.8}$ M$_\odot$ (2$\sigma$) \citep{gilman_warm_2020}. Improvements in this measurement can be made by increasing the sample of lenses, improving the lens modeling techniques applied to interpret the data, improving flux-ratio measurement sensitivity, and choosing sources with intrinsically smaller sizes. 

Quasar warm dust serves as an attractive light source for flux-ratio anomaly measurements. This dust component has temperatures of hundreds of Kelvin and dominates the quasar spectral energy distribution at rest-frame wavelengths of $\sim 8-12\mu$m. It has typical sizes of $\sim 0.1-10$ pc \citep{burtscher_diversity_2013, leftley_parsec-scale_2019}, with minimal scaling with quasar luminosity. This is much smaller than the nuclear narrow-line emission with FWHM$\sim$100 pc \citep{muller-sanchez_outflows_2011, nierenberg_detection_2014, nierenberg_probing_2017}. Figure~\ref{fig:diff_mag} demonstrates an example of the magnification induced by a perturbing subhalo on a source with a characteristic size scale of the narrow-line emission compared with the warm dust emission. The size of the quasar warm dust emission region is excellent for dark matter studies, as it is large enough to be unaffected by microlensing while still being small enough to be significantly magnified by individual low-mass dark matter halos. It is also bright and ubiquitous.

Quasar warm dust has long been recognized as a potential source for analyses of dark matter through strong lensing. Several studies have undertaken IR studies of strongly lensed quasars out to \emph{observed frame} 
10~$\mu$m \citep{agol_keck_2000, chiba_subaru_2005,fadely_near-infrared_2011, jones_image_2019, 
macleod_detection_2009, macleod_detection_2013,ross_uv-mid-ir_2009}.  \citet{chiba_subaru_2005} and \citet{macleod_detection_2009} both measured flux ratios to be consistent with results from lensed radio jets. 
These studies probed \emph{rest-frame} wavelengths of $\sim$3-5 $\mu$m, where we expect light from the quasar accretion disk as well as both hot and warm dust components \citep[see e.g.][and references therein]{stalevski_gravitational_2012, sluse_mid-infrared_2013}. Measuring the flux ratios at even redder wavelengths, where the warm dust dominates the SED, may provide an even more robust constraint of dark matter structure. This has now become possible with JWST, which has both the spatial resolution and sensitivity to measure lensed quasar flux ratios to \emph{rest-frame} 8 $\mu$m given typical source redshifts. 

Here we introduce our survey JWST-GO-2056 (PI: Nierenberg) of 31 quadruply lensed quasars in which we use multi-band Mid-Infrared Instrument (MIRI) imaging with JWST to measure the warm dust flux ratios. Given typical source sizes of $1-10$ pc, and target flux ratio precision of $3\%$, dark matter halos with masses below 10$^7$ M$_\odot$ can cause a significant perturbation to the flux ratios. No existing dataset has demonstrated the capability to reveal the presence of dark halos on these scales across cosmological distances. Detecting a population of halos at 
$10^7$M$_{\odot}$ would have profound consequences for dark matter physics. Independent confirmation of the presence of dark halos through lensing would verify a key prediction of the $\Lambda \rm{CDM}$ paradigm, complementing other probes of low-mass dark matter structure, such as studies of dwarf galaxies \citep[e.g.][]{nadler_constraints_2021,dekker_warm_2022,slone_orbital_2023} and stellar streams \citep{bovy_linear_2017,banik_evidence_2021}. Non-detection of these low-mass halos would falsify CDM, and an inference of their central density profiles and concentrations would improve existing bounds from lensing on self-interacting dark matter, fuzzy dark matter, and the matter power spectrum \citep[see][and references therein]{vegetti_strong_2023}. 

This paper is organised as follows. In Section \ref{sec:survey_design}, we describe the survey design and sample selection. In Section \ref{sec:0405data} we present measurements for the first target observed for our program, DES J0405-3308 \citep{anguita_strong_2018}. In Section \ref{sec:model}, we describe how we measure the light components. In Section \ref{sec:sed_fitting} we present our model for fitting the quasar spectral energy distribution. In Section \ref{sec:discussion}, we discuss our results in light of previous measurements of this system. In section \ref{sec:forecast} we estimate our sensitivity to dark matter halos for the full survey. In Section \ref{sec:summary} we provide a summary of the major conclusions of this paper. 
In order to calculate physical sizes, we assume a flat $\Lambda$CDM cosmology with
$h=0.7$ and $\Omega_{\rm m}=0.3$. 

\section{The quasar mid-IR spectral energy distribution, and survey design}
\label{sec:survey_design}

The goal of this program is to measure the flux ratios of strongly lensed warm dust emission of quasars in order to constrain the properties of dark matter.  
Quadruply imaged quasars were selected from the current known sample of $\sim$50 systems \citep{inada_sloan_2012,lemon_gravitationally_2017, agnello_meets_2018,agnello_quasar_2019, delchambre_gaia_2019,lemon_gravitationally_2019,stern_gaia_2021}. These systems were discovered through a combination of data from wide-field surveys including the Sloan Digital Sky Survey \citep{york_sloan_2000}, the Panoramic Survey Telescope and Rapid Response System \citep{chambers_pan-starrs1_2016}, Gaia \citep{gaia_collaboration_gaia_2023}, the Wide-field Infrared Survey Explorer \citep{wright_wide-field_2010}, and the Dark Energy Survey \citep{dark_energy_survey_collaboration_dark_2016}. We first describe the properties of the quasar mid-infrared spectral energy distribution that are relevant to our measurement and explain how this impacted our observation strategy and lens selection. After selecting based on the criteria outlined in the following subsections, the final sample contains 31 lenses.
We will provide detailed information for each target in the papers that present flux ratios for those targets.

\subsection{Photometric requirements for spectral energy distribution fitting}
\label{sec:sed}

Our goal is to isolate emission coming from physical regions more extended than $\sim0.1$ pc in order to ensure that these regions subtend an angular size of $\sim$mas,
and are therefore not contaminated by stellar microlensing in the lens galaxy. This in turn ensures that the flux ratios we measure are sensitive only to the presence of low-mass dark matter halos rather than stellar microlensing or intrinsic variability.

The current picture of the mid-IR emitting region of quasars has been built up using a combination of narrow-band imaging, reverberation mapping, and high-resolution interferometric measurements. One model is consistent with all of these observations. In this model, the mid-IR SED of quasars is composed of three relatively distinct sources of emission. At wavelengths below 2 microns, there is significant emission from the quasar accretion disk, which has physical scales of light-days \citep[e.g.][]{wambsganss_interpretation_1990, wanders_steps_1997,anguita_multiple_2008,fausnaugh_space_2016}, corresponding to angular sizes of $\mu$as at typical source redshifts. 
At redder wavelengths, the spectral energy distribution becomes dominated by a `hot' dust region with peak flux emitted at temperatures ranging from 1000--1400 K ($\sim$3 $\mu$m) \citep[e.g.][]{bosman_first_2023}. This emission is associated with dust near the sublimation temperature that marks the inner boundary of the dusty region of the quasar and has characteristic size scales of order 0.05--0.2 pc \citep{suganuma_reverberation_2004, suganuma_reverberation_2006, mor_hot-dust_2011,gravity_collaboration_image_2020}, depending on quasar luminosity. In addition to this, there is a `warm' dust component \citep[see e.g.][and references therein]{honig_redefining_2019}, which dominates the SED at wavelengths of 8--12 $\mu$m. This component is observed to subtend scales of $\sim 0.1-10$ pc, with little or no scaling with luminosity \citep{burtscher_diversity_2013, leftley_parsec-scale_2019}. 

\begin{figure}
\centering
\includegraphics[width=0.8\linewidth,trim=0cm 0cm 0cm 0cm,clip]{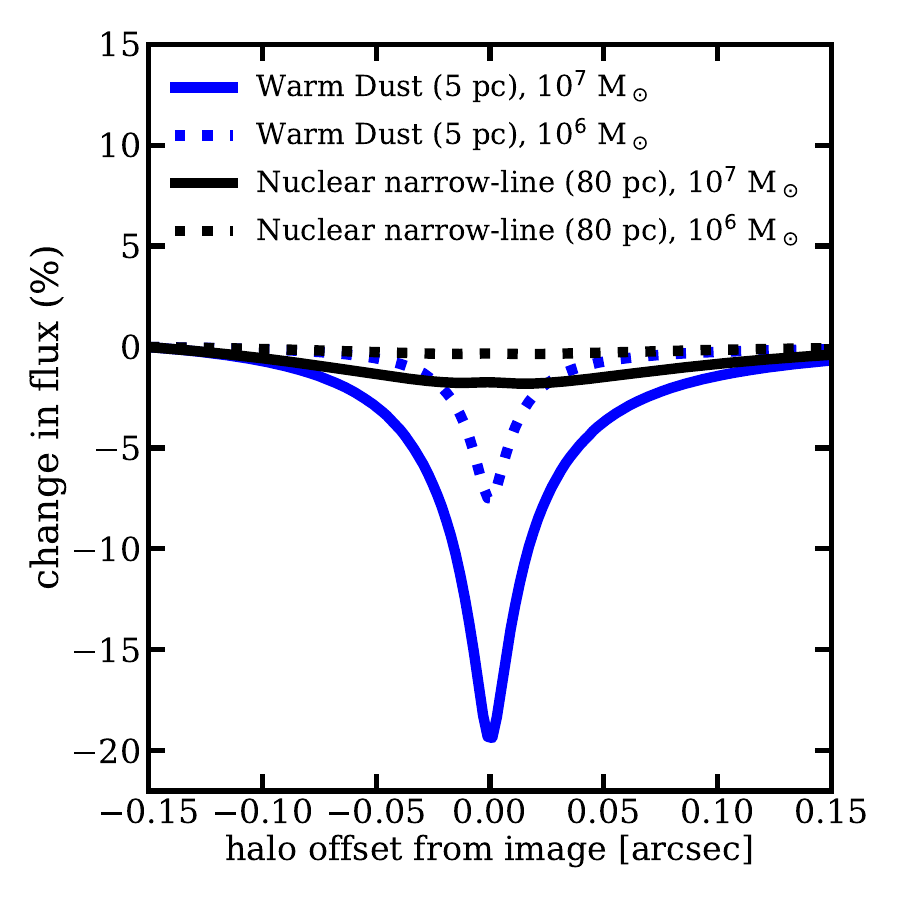}
\hspace{-5mm}
\caption{Illustration of the differential magnification of a saddle image of a quadruply imaged quasar with a Gaussian light distribution by a perturbing NFW subhalo with a mass of 10$^6$ (dashed lines) and 10$^7$~M$_\odot$ (solid lines), as a function of the position of the subhalo relative to the center of the lensed image. Per cent differences in flux are relative to a model without a subhalo. The subhalo significantly alters the flux of the smaller source (blue lines) with FWHM typical of the quasar warm dust region, but it is not massive enough to significantly affect the larger source (black lines) with FWHM typical of the quasar nuclear narrow-line region. The JWST program described in this work aims to have sensitivity to the effects of 10$^7$ M$_\odot$ subhalos, which are not expected to contain detectable gas or stars. Our final measurements will be made statistically by generating populations of dark matter halos both in the lens and along the line of sight, and by marginalizing over uncertainties in the deflector macromodel and source properties as described in \citep{gilman_probing_2019, gilman_warm_2020}. \label{fig:diff_mag}
}
\end{figure}

The size of the warm torus makes it both insensitive to microlensing, as well as relatively more sensitive to low mass perturbations than the larger narrow-line region used in previous flux-ratio anomaly studies \citep{nierenberg_double_2020, gilman_warm_2020}. Figure \ref{fig:diff_mag} illustrates this for the case of a saddle image in a quadruply imaged quasar. Saddle images are located at a saddle point in the time-delay surface of the lens and are therefore particularly sensitive to the effects of small-scale perturbations. The smaller source with FWHM of 5 pc, characteristic of the quasar warm dust emitting region, is significantly more perturbed by the subhalo than is the more extended source with FWHM of 80 pc, characteristic of the quasar nuclear narrow-line region. We are aiming for measurements that are sensitive to the presence of individual 10$^7$ M$_\odot$ NFW halos \footnote{In CDM, we expect large numbers of such subhalos and therefore we will model their collective effects.}. We selected this mass target as it is below the threshold at which the majority of halos are believed to contain detectable galaxies \citep[e.g][]{nadler_dark_2021}. Based on these simulations, we aim for a target flux ratio measurement signal to noise of 3\%.

\citet{sluse_mid-infrared_2013} performed microlensing analyses of simulated lensed quasars spectral energy distributions and demonstrated that lensed quasar images could be significantly affected by microlensing at rest-frame wavelengths blue-ward of 8 $\mu$m, because of the small physical size of the hot dust emitting region, and the quasar accretion disk. Therefore, ideally, a flux-ratio study of quasars would probe only the warm dust emission at rest-frame wavelengths beyond 10 $\mu$m and redder in order to avoid contamination. The reddest MIRI imaging filter is 25.5 $\mu$m. Such a restriction on rest-frame wavelength would enable us to study only lensed quasars with redshifts below 1.5. 

In order to expand our sample to higher source redshifts, and to ensure a lack of microlensing contamination at lower redshifts, we use multi-band imaging spanning the near-to-mid-IR SED of the quasar to constrain the relative contributions of the quasar accretion disk and the hot and warm dust for each lensed image. Based on simulations presented in a companion paper (Sluse et al., in prep.), such multi-band imaging enables the identification of lensed images affected by significant microlensing and can be used to reduce systematic uncertainties relative to single-band imaging only.

We adopted the following strategy to measure the spectral energy distribution of lensed quasar images. For all lenses, we obtained imaging in F560W, F1280W, and F1800W to obtain a constraint on the relative brightness of the quasar accretion disk and hot dust emission. We also required the reddest filter to measure rest-frame 6 $\mu$m or redder.  Thus, for quasars with redshifts $z>2$, we required the faintest lensed image to be detectable in F2550W.  Our target signal-to-noise was 100. Using the pre-launch JWST Exposure Time Calculator, this corresponded to a minimum lensed image flux of 1 mJy. The faintest lensed image fluxes were estimated by applying the optical flux ratios by the unresolved total flux measured in WISE W4 (22.4 $\mu$m). 

For source quasars with redshifts $z<2$, F2100W (rest-frame 8 $\mu$m or redder) provides sufficiently red wavelength coverage to mitigate microlensing. This filter is much more sensitive than F2550W given the lower background and more compact point-spread function (PSF), and thus we did not impose a minimum flux requirement for these targets beyond an unresolved detection of the lens in W4 (total W4 flux for all four images greater than $\sim 3$ mJy). 

Given typical quasar SEDs, and the sensitivity of MIRI imaging as a function of wavelength, these criteria were sufficient to ensure that the quasar flux ratios could be measured with adequate signal-to-noise in the three bluer filters.

In addition to the sensitivity requirements, we selected lenses with a minimum image separation of 0\farcs1 for accurate image deblending, given that the highest resolution imaging is in F560W with a PSF FWHM of 0\farcs2.

\subsection{Macromodel Requirements}
Lenses were selected to have four images to constrain the smooth mass distribution, which is used as a baseline for flux ratio anomaly studies. Furthermore, we required that the lens have a `simple' deflector light distribution with no significant disk, and only a single massive deflector was needed to reproduce the observed image positions. 

\section{Observations and initial reduction}
\label{sec:0405data}

The first system to be observed was DESJ040559.7-330851.00 \citep{anguita_strong_2018}. This lens has source redshift of $z_s=1.713$ and a photometrically estimated deflector redshift of $z_d\sim 0.3$ \citep{gilman_warm_2020}. 
DESJ0405-3308 has an unresolved W4 flux of 7.7 mJy. Assuming the optical flux ratios are identical to the F2550W flux ratios, this would indicate an expected faint image flux of approximately 1.3 mJy. Based on our photometric criteria, this was bright enough to use F2550W as the reddest filter for this target, enabling us to measure fluxes at rest-frame $\sim$9.4 $\mu$m, where we expect little to no contamination from microlensing. For this system, the spectral energy distribution will provide a useful test of our SED fitting method.

Observations for DESJ0405-3308 were obtained on October 27, 2022.  Exposure times were 58 s in F560W, F1280W, and F1800W and 574 s in F2550W. All exposures were divided into a three-point dither pattern to improve spatial resolution and mitigate cosmic rays.

Initial calibration was performed using the default JWST data calibration pipeline\footnote{Using the {\tt jwst\_1041.pmap} context file.} \citep{greenfield_calibration_2016, bushouse_jwst_2022}.  Sky subtraction of Level 2 data products was performed using customized routines\footnote{Based on \url{https://github.com/STScI-MIRI/Imaging_ExampleNB}} before drizzling to produce the final images.  The final pixel scale was set to 0\farcs11 per pixel, identical to the native detector pixel scale. Reduced images in each filter are shown in Figures 3-6.
\footnote{The MIRI imager sensitivity at long wavelengths appears to have dropped over the first six months of its operation: https://www.stsci.edu/contents/news/jwst/2023/miri-imager-reduced-count-rate?Type=miri.
If we simply assume that the time history of this sensitivity drop was similar to what has been more definitively measured for MIRI spectroscopy at $\lambda \ge 20 \mu $m --
https://jwst-docs.stsci.edu/jwst-calibration-pipeline-caveats/jwst-miri-mrs-pipeline-caveats,
this would imply that the 2550W fluxes we measure are underestimated by about 10 per cent.}

\section{Image Flux Measurement}\label{sec:model}
Our goal was to accurately measure the lensed image fluxes in the presence of other light components including the lensed quasar host galaxy (which appears as a lensed arc) and the deflector galaxy light. We adopted a forward modelling approach to measure the lensed quasar image fluxes in all four filters.  The model consisted of a combination of up to four light components depending on the filter, as described below.

{\bf Lensed quasar images: } The quasar light is dominated by the accretion disk and hot and warm dust on angular size scales of micro- to milli-arcseconds. Given that this is smaller than the smallest imaging PSF with FWHM of F560W of 0\farcs2, we treated these components as point sources. We wished our measurement to have as little dependence as possible on the gravitational lensing model, as the image fluxes will later be used to constrain this model with dark matter substructure. Therefore, we did not associate the point source fluxes or positions with a lens model but rather treated them as completely independent. This is the same procedure one might adopt if, for example, there were foreground stars in the data. 

{\bf Deflector light distribution:} The lens galaxy is detected in F560W and F1280W. We modelled this light distribution as an elliptical S\'ersic profile \citep{sersic_influence_1963}.

{\bf Lensed quasar host galaxy:} The lensed host galaxy of the quasar is apparent as an extended arc in F560W, F1280W and F1800W. We modelled the unlensed quasar host galaxy light distribution as an elliptical S\'ersic profile. To produce the observed gravitationally lensed arc, we included a gravitational lensing model for the deflector mass distribution.  We adopted an elliptical power-law model \citep{tessore_elliptical_2015}, with external shear.

\subsection{Point spread function fitting \label{sec:psf_fitting}}

We used \webbpsf \footnote{Development branch 1.2.1.} \citep{perrin_simulating_2012, perrin_updated_2014-1} to fit the PSF in our data. We used a super-sampling of 3 in order to enable improved astrometric precision, and because of the large detector pixel scale relative to the sizes of the light features such as the lensed quasar host galaxy. At the time of writing, this software was in active development to update the models to match observed optics and detector properties.  The default parameters provided a poor fit to the observed data due to detector-level effects. The dominant discrepancy was due to inter-pixel capacitance and charge diffusion in the detector \citep[e.g.][]{argyriou_brighter-fatter_2023}. A preliminary model for the charge diffusion effect has been implemented. However, at the time of writing, this was only in the detector-sampled PSF models, while we required a super-sampled PSF model given the large pixel size relative to the light features.

As an alternative, we found that the PSF could be modelled by varying the \webbpsf Gaussian `{\tt jitter\_sigma}' parameter. The `jitter' effect is implemented in \webbpsf by convolving the PSF model with a Gaussian kernel to account for spacecraft motion. In practice, the jitter effect has a nearly equivalent impact on the data, as does charge diffusion.\footnote{M. Perrin: Private communication.} The {\tt jitter\_sigma} value was optimized for each filter as described in the following. 

We used blackbodies at the redshift of the quasar to account for the wavelength dependence of the PSF. The temperature of each blackbody was optimized separately for each filter.
Although in principle the PSF spectrum should be connected to the SED of the quasar (rather than a single blackbody), we found that a single blackbody model for the PSF source provided an excellent fit to the data. We defer incorporating additional complexity in the PSF simulation until the PSF model has been further refined based on in-flight results.

In addition to the charge diffusion effects, F560W displays a prominent `cruciform' artifact \citep{gaspar_quantum_2021, wright_mid-infrared_2023}, which is a wavelength-dependent, detector-level artifact apparent beyond the first Airy ring.  The second extension output of \webbpsf provides a model for this feature that provides an improved fit relative to the PSF model without it. However, residuals owing to the cross artifact were still prominent in our data. We therefore fit the F560W data in a relatively small region where the cross feature was sub-dominant.

\subsection{Modelling Procedure}

We adopted an iterative approach to fitting our images, switching between optimizing the PSF parameters ({\tt jitter\_sigma} and blackbody temperature), and the parameters associated with  light sources and gravitational lensing model  until both inferences were returning stable results.  Due to the small number of stars in the field of view, and their very different SED from our quasar images, we fit the PSF parameters directly using our lensed quasar images. 

We fit the three images that contain the lensed quasar host galaxy simultaneously. We required the image positions, gravitational lens model, and the centroids of the deflector and source light to be the same between the three filters but allowed all other model parameters to vary between the three filters. F2550W, which contained only four point sources, was fit independently with no lens model and only the four independent PSFs.

After finding the best-fitting model parameters, uncertainties were estimated using a Markov Chain Monte Carlo with the PSF held fixed at the best-fitting value obtained from the previous steps. Given that the flux ratios show no variation over a broad range of PSF model parameters (including those that provide a poor overall fit to the data), we do not anticipate that this choice will make a significant impact on the estimate of the flux-ratio uncertainties. We used {\tt lenstronomy} \citep{birrer_gravitational_2015,birrer_lenstronomy_2018,birrer_lenstronomy_2021} for all image fitting and simulation.

\subsection{Results of forward modelling and uncertainty estimation}

\begin{figure}
\centering
\includegraphics[width=\linewidth]{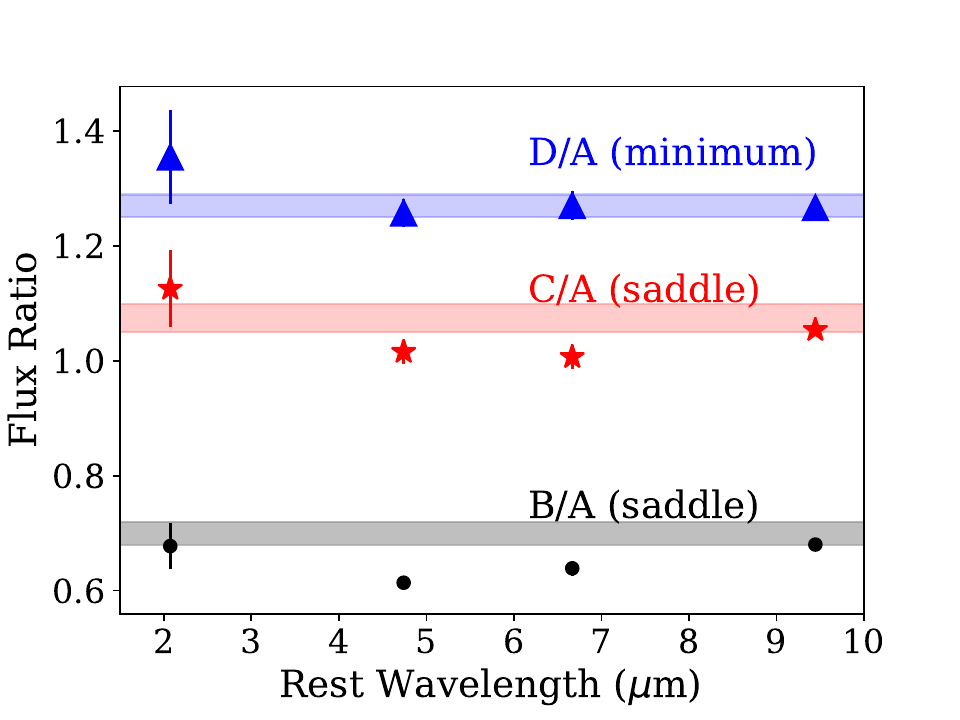}
\hspace{-5mm}
\caption{The measured flux ratios with respect to image A as a function of rest frame wavelength. Colored bands indicate the 68\% confidence interval of the corresponding warm torus component, which is not expected to be microlensed.  The labels indicate whether the image is located at a minimum or saddle point of the time delay surface. Image A is a minimum. Rest wavelengths blue-ward of 8 $\mu$m rest frame have significant contributions from the hot dust and accretion disk that are small enough to be microlensed by stars in the lens galaxy and/or time variable on the day-to-week time scales. \label{fig:fr_lam}}
\end{figure}

The best-fit PSF parameters are given in Table \ref{tab:psfPars}, 
and the measured image fluxes and positions are given in Table~\ref{tab:fluxes}. Figure \ref{fig:fr_lam} shows the measured flux ratios as a function of wavelength.

We do not report the lens model parameters. Owing to the limitations of the current PSF model as well as the fact that the quasar images are treated as independent foreground objects, the lens and light model parameters we infer cannot be meaningfully compared to other studies for this system, which were based on Hubble Space Telescope data with a well-modelled PSF \citep{shajib_is_2019, schmidt_strides_2023}. Ultimately, for our gravitational lensing dark matter measurement, we will apply the approach used by \cite{gilman_probing_2019, gilman_warm_2020} in which only the image positions and flux ratios are used to constrain the mass distribution of the deflector. This allows for a high degree of flexibility in the smooth mass distribution used as the baseline for the flux-ratio comparison \citep[see also][]{nierenberg_double_2020}. Below we discuss our tests for the dependence of measurement uncertainty on model choices.

The formal statistical uncertainties for the image fluxes, positions, and flux ratios were extremely small. Here we describe how we estimated systematic uncertainties due to model choices. When estimating uncertainties, it is important to make the distinction between \emph{absolute fluxes}, which are relevant to SED fitting described in Section \ref{sec:sed_fitting}, and \emph{flux ratios}, which are the key quantity for gravitational lensing estimates. 

{\bf Position Uncertainties:} We estimate the systematic uncertainties by comparing the measured relative image positions with those measured in HST WFC3-IR F140W direct imaging from \cite{nierenberg_double_2020}, and find maximum relative offsets of 0\farcs007 in the lensed image positions. This is much smaller than the pixel sizes of 0\farcs11 for JWST MIRI and 0\farcs13 for HST WFC3-IR. 

{\bf Light component modelling:} We performed several tests of the systematic uncertainties on measured image fluxes and flux ratios.  These included: 1) Fitting the light in the imaging bands together and requiring the model light components to have the same parameters except amplitude in all three bands; 2) Performing the fits in the three filters separately and allowing the lens model to be different in each filter; 3) Restricting the source light to be round in shape; 4) Restricting the host mass profile to have a slope of $\gamma_{\rm{p}} = 2$ rather than allowing it to vary freely; and 5) Fixing the image positions to those specified by the lens model, rather than treating them as completely independent foreground light sources. As an additional test on the flux ratios, we measured the flux ratios before and after including the lensed quasar host galaxy. 

The extended source was most significant in F560W, contributing approximately 40\% of the flux at the location of the quasar images. In F1280W and F1800W the flux was less than 10\% at the location of the quasar images. This is reflected in the systematic uncertainties from the tests above, in which we found that the absolute fluxes varied by 5\% in F560W and F1280W and 2\% in F1800W, and the flux ratios varied by up to 6\% in F560W and 1\% in F1280W and F1800W.

{\bf PSF uncertainties:} We found that variations in the choice of PSF model impacted the \emph{absolute} image fluxes by $10\%$ or less. We also tested for variation of PSF within a filter as a function of image brightness. Therefore we did an additional fit of the F2550W data, allowing each point source to have a different {\tt jitter\_sigma} value. We found no significant variation in the value of this parameter between the four images, indicating that the adoption of a single PSF model was sufficient for this system. Furthermore, even with the variable PSF, the flux ratios and fluxes varied by less than 1\% relative to a fit in which the PSF was the same for all four images.

{\bf Instrument Calibration:}
The absolute flux calibration uncertainties for MIRI have not been estimated at the time of writing. In August 2023 a significant wavelength-dependent loss in sensitivity of 3\% for F1280W, 8\% for F1800W and 18\% for F2550W was reported for the MIRI imager relative to the commissioning sensitivity measured in Summer 2022.\footnote{\url{https://www.stsci.edu/contents/news/jwst/2023/miri-imager-reduced-count-rate?page=1&keyword=MIRI}}  The sensitivity loss seems to have occurred over time. At the time of writing it is not known what the sensitivity loss was at the time of the observations for this program (October 2022), therefore we include the August 2023 reported loss values as an additional systematic uncertainty in our absolute flux measurements.

{\bf Conclusion of uncertainty estimate testing:}
Based on our tests of systematic sources of uncertainty, we find that the absolute flux uncertainty is likely dominated by the uncertainty in the instrument calibration. For this work, we adopt 15\% flux uncertainties in F560W, F1280W, and F1800W, and 20\% flux uncertainties in F2550W based on our current knowledge of the detector calibration. We expect these uncertainties to become smaller in the near future as the instrument behavior is better understood.

The dominant source of flux ratio uncertainty in F560W was 6\% from modelling the lensed quasar host galaxy, while the uncertainties related to PSF modelling and lensed quasar host galaxy modelling were comparable for the flux ratio measurements in F1280W and F1800W. We adopt flux ratio uncertainties for 2\% in these filters. For F2550W, which did not have an apparent lensed quasar host galaxy, we estimate 1\% flux ratio uncertainties.

\begin{figure*}

    \includegraphics[width=\linewidth,trim=0.5cm 0.5cm 0.5cm 0.5cm,clip]{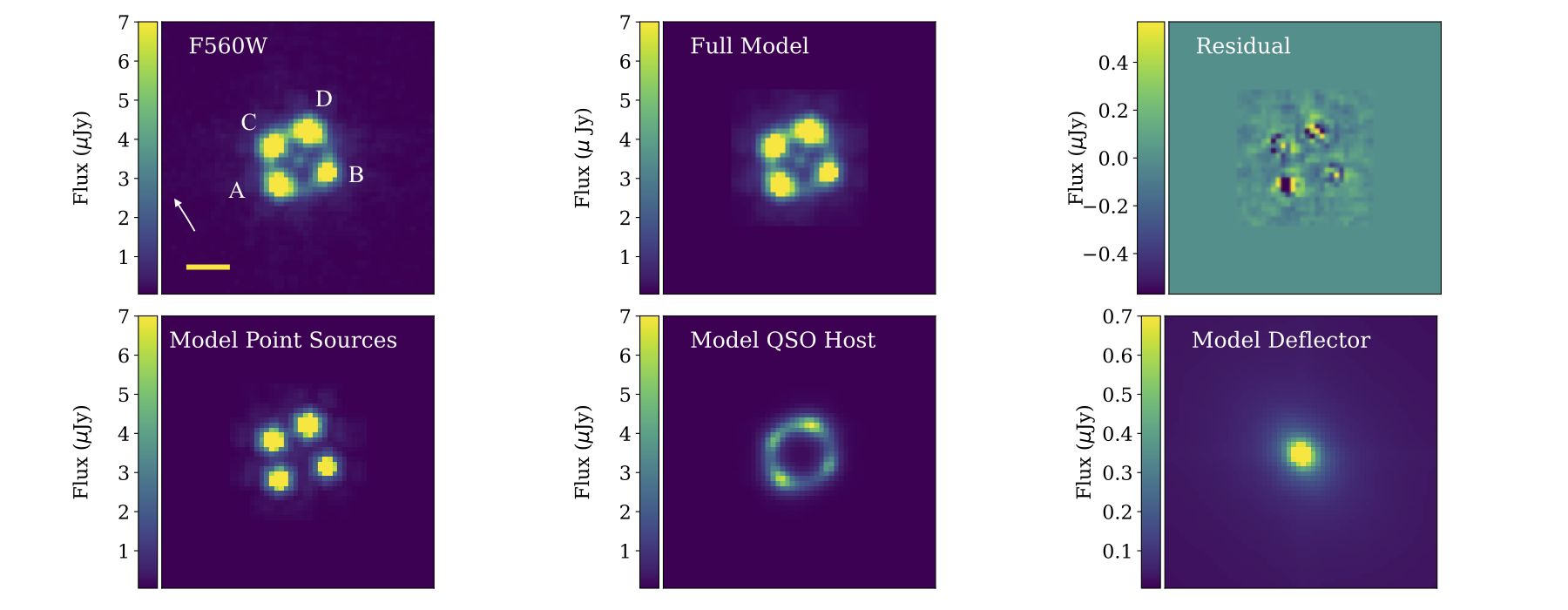}
\hspace{-5mm}
\caption{ \emph{Upper panels}: From left to right, comparison of original F560W image, best-fit model, and residuals. \emph{Lower panels}: Separate light model components. From left to right, model point sources, lensed quasar host galaxy, and deflector light distribution. The yellow bar in the lower left of the data image indicates 1 arcsecond. The arrow indicates North. 
}
\end{figure*}

\begin{figure*}

\includegraphics[width=\linewidth,trim=0.5cm 0.5cm 0.5cm 0.1cm,clip]{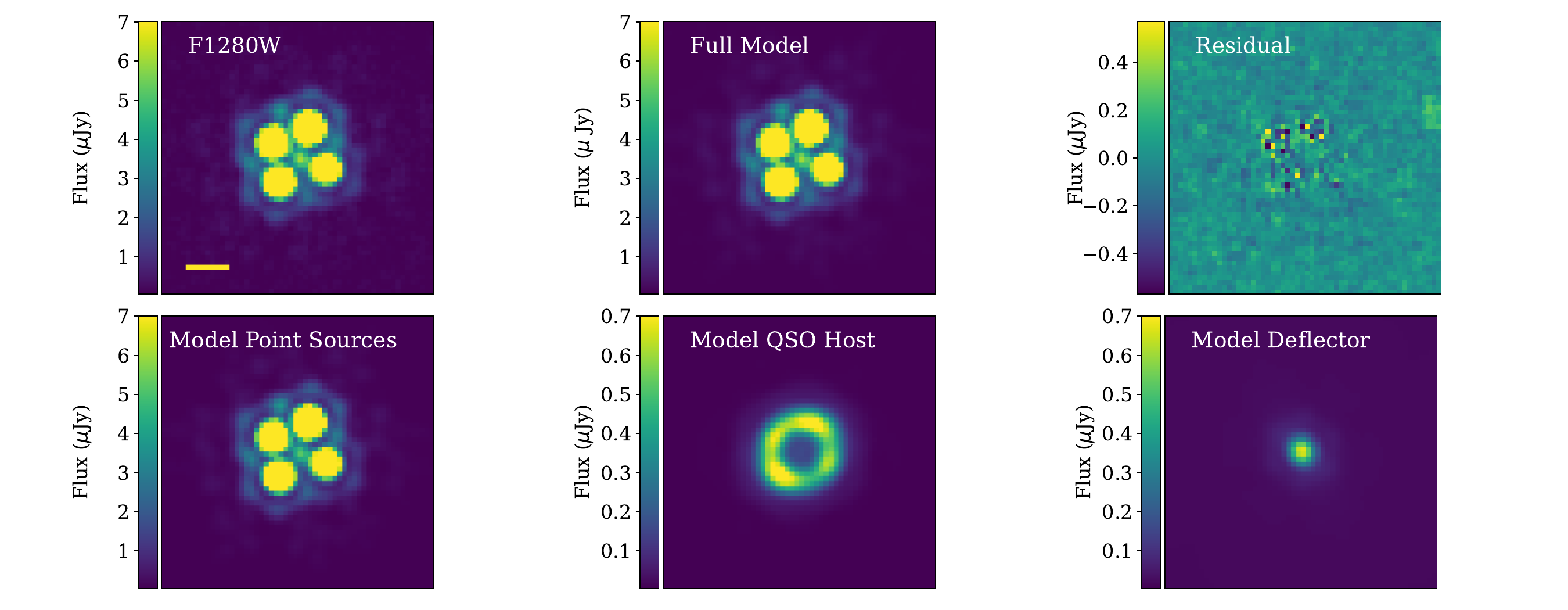}
\hspace{-5mm}
\caption{ \emph{Upper panels}: From left to right, comparison of original F1280W image, best-fit model, and residuals. \emph{Lower panels}: Separate light model components. From left to right, model point sources, lensed quasar host galaxy, and deflector light distribution. The yellow bar in the lower left of the data image indicates 1 arcsecond. 
}
\end{figure*}

\begin{figure*}

\includegraphics[width=\linewidth,trim=0.5cm 0.5cm 0.5cm 0.5cm,clip]{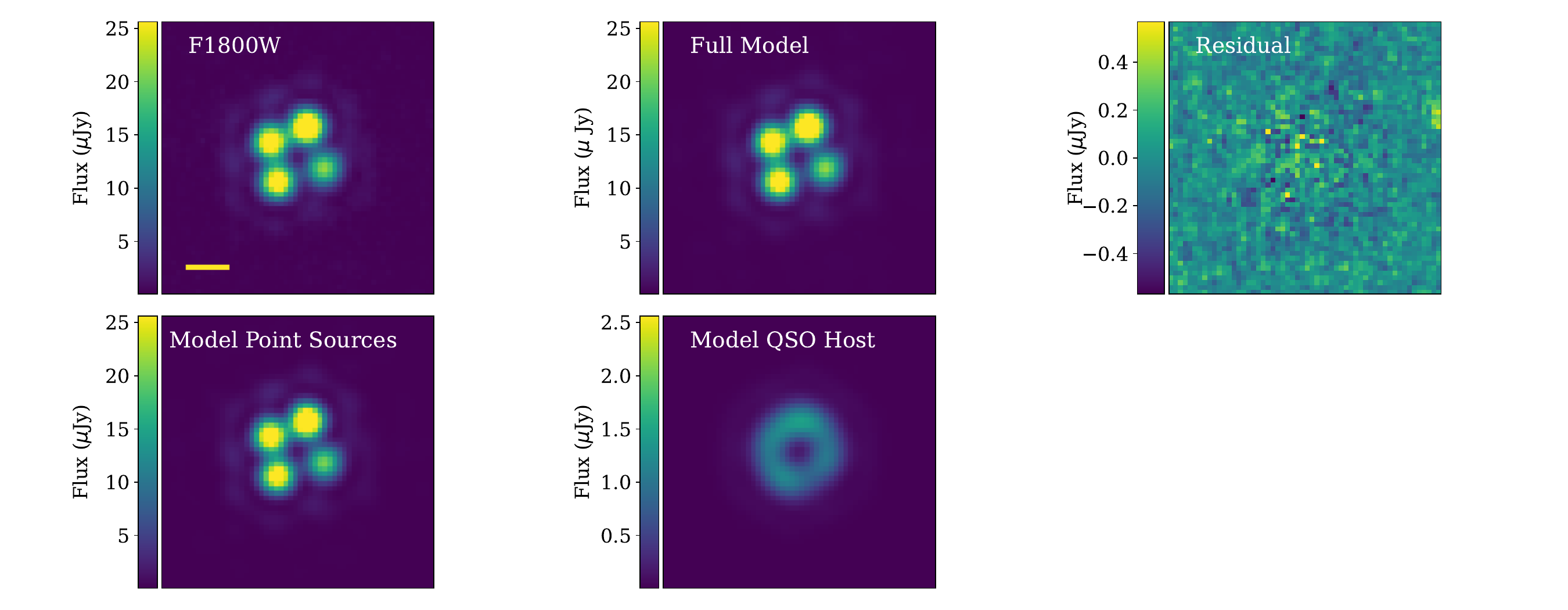}
\hspace{-5mm}
\caption{ \emph{Upper panels}: From left to right, comparison of original F1800W image, best-fit model, and residuals. \emph{Lower panels}: Separate light model components. From left to right, model point sources and lensed quasar host galaxy. The deflector light is not detected in this filter, thus it is not included in the model. The yellow bar in the lower left of the data image indicates 1 arcsecond.  
}
\end{figure*}

\begin{figure*}
\centering
\includegraphics[width=\linewidth,trim=0.5cm 2cm 0.5cm 1cm,clip]{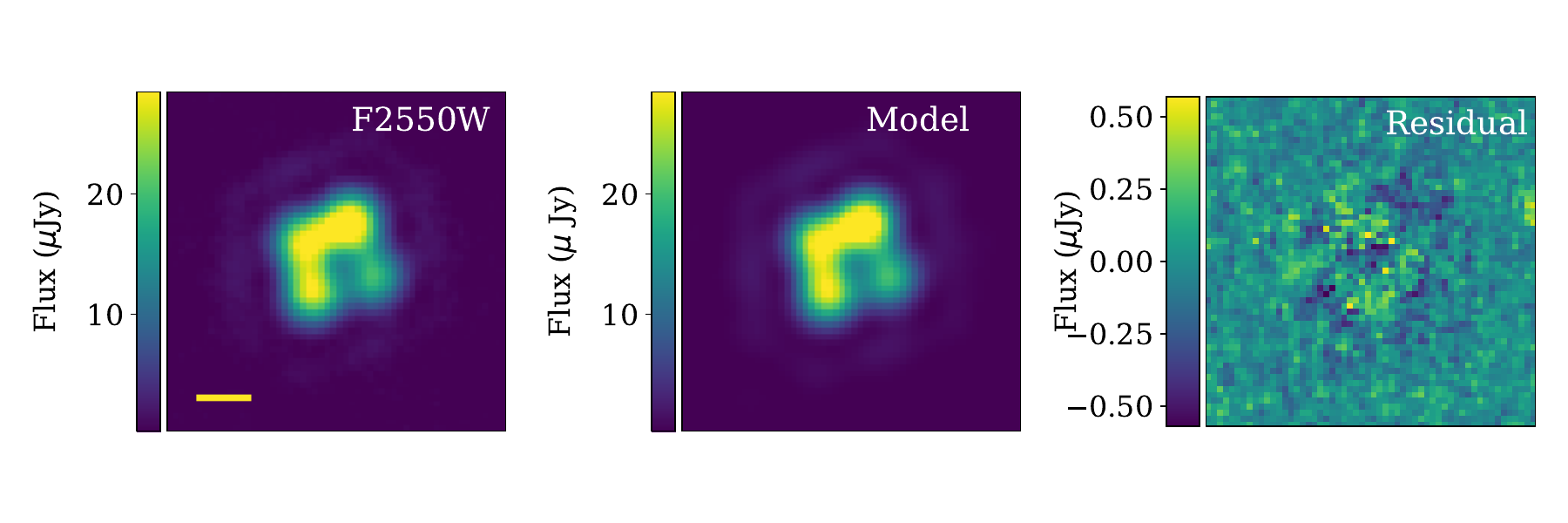}
\caption{ From left to right, comparison of original image, model, and residuals. The light model in F2550W consists of only the point source contribution. The yellow bar in the lower left of the data image indicates 1 arcsecond.}

\end{figure*}

\begin{table*}
\small
\centering
\begin{tabular}{lllllllll}
\hline
\hline
Parameter &  F560W & F1280W & F1800W & F2550W  \\
\hline
{\tt jitter\_sigma}   &  0\farcs063                 & 0\farcs061    & 0\farcs073   & 0\farcs075  \\
Temperature (K)  &  1120        & 700 & 680  & 250   \\
\hline
\end{tabular}
\caption{Best fit \webbpsf {\tt jitter\_sigma} and blackbody temperature for each filter. These values were inferred for our data using \webbpsf development version 1.2.1, and do not include inter-pixel capacitance effects. \label{tab:psfPars}}
\end{table*}

\begin{table*}
\small
\centering
\begin{tabular}{lllllllll}
\hline
\hline
Image & dRa & dDec &  F560W & F1280W & F1800W & F2550W  \\
\hline
A   & 1.065         & 0.318        & 0.396                 & 1.06     & 1.38   & 2.647   \\
B   & 0             & 0            & 0.279                 & 0.656     & 0.875   & 1.787   \\
C   & 0.721         & 1.152        & 0.459                 & 1.08     & 1.42   & 2.790   \\
D   & -0.153        & 1.018        & 0.536                 & 1.34     & 1.73   & 3.357   \\
\hline
\end{tabular}
\caption{
Measured image positions and fluxes in units of mJy.  Image positions are measured from the F2550W data. Image naming follows \protect\cite{shajib_is_2019}, and image labels are shown in Figure 3. We estimate the flux ratio (absolute flux) uncertainties to be 6\% (15\%), 2\% (15\%), 2\% (15\%), 1\% (20\%) in F560W, F1280W, F1800W, and F2550W respectively. The Right Ascension and Declination offsets of the quasar images with respect to image B are within 0\farcs007 of those measured by \protect\cite{nierenberg_double_2020}.}
\label{tab:fluxes}
\end{table*}

\section{SED Fitting }
\label{sec:sed_fitting}
In this section, we describe how we used the MIRI four-band photometry to fit the multi-component SED and isolate light coming from the warm dust region of the quasar, which is extended enough to avoid contamination from micro-lensing as described in Section~\ref{sec:survey_design}.

We followed \citet{sluse_mid-infrared_2013} and adopted a simple three-component model of the quasar spectral energy distribution. This is composed of power-law continuum emission from the quasar accretion disk combined with two blackbodies representing the hot dust component, with prior temperature range of 500--1800 K, and the warm dust component, with prior temperature range 100--500 K. We did not include emission lines such as PAH emission, which we expect to make a small contribution to the broad-band fluxes \citep[e.g.][]{garcia-bernete_high_2022}.

Our SED model allowed for independent variation of each component amplitude for each lensed image  
to account for the fact that both the quasar accretion disk and the hot dust are small enough to be affected by microlensing. This also accounts for intrinsic flux variation of the accretion disc on timescales shorter than the time delay between the lensed images (of order days) \citep{schmidt_strides_2023}. 
We performed the SED fit simultaneously for all four images. The temperatures of the hot and warm dust blackbodies were allowed to vary as free parameters but were restricted to be the same for all images. The overall SED amplitudes were also allowed to vary independently to account for different overall magnifications for the lensed images. 

 When fitting the lensed quasar SEDs, we computed the joint likelihood that each set of model parameters would reproduce the observed \emph{flux ratios} (B/A, C/A, and D/A) in each filter as well as the likelihood that the model matched the \emph{absolute fluxes} for image A in each filter. Model SEDs were transformed into band fluxes following \citet{gordon_james_2022}\footnote{\footurl}. We used {\tt emcee} \citep{foreman-mackey_emcee_2013} to estimate the posterior probability distribution. 
 
\begin{figure*}
\centering
\includegraphics[width=0.45\linewidth,trim=0cm 0cm 0cm 0cm,clip]{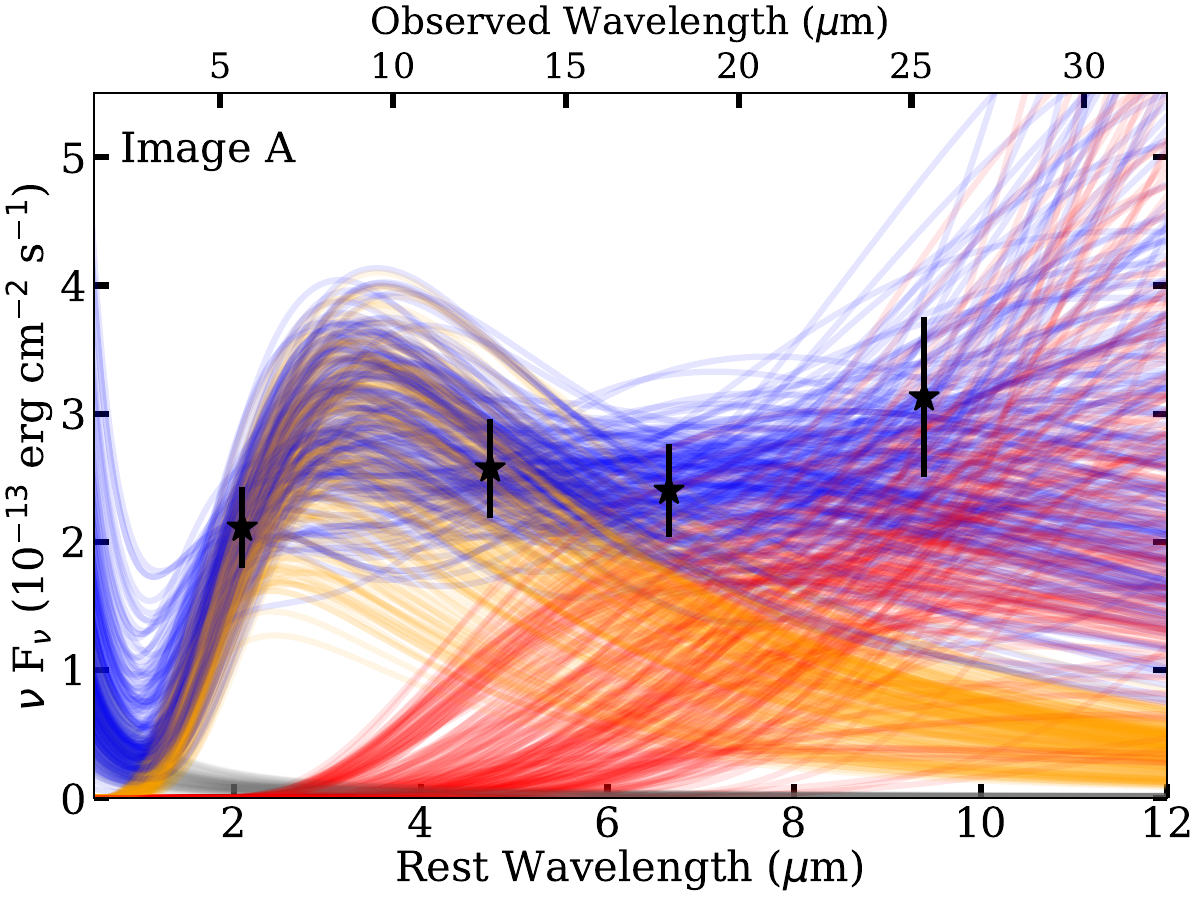}
\includegraphics[width=0.45\linewidth,trim=0cm 0cm 0cm 0cm,clip]{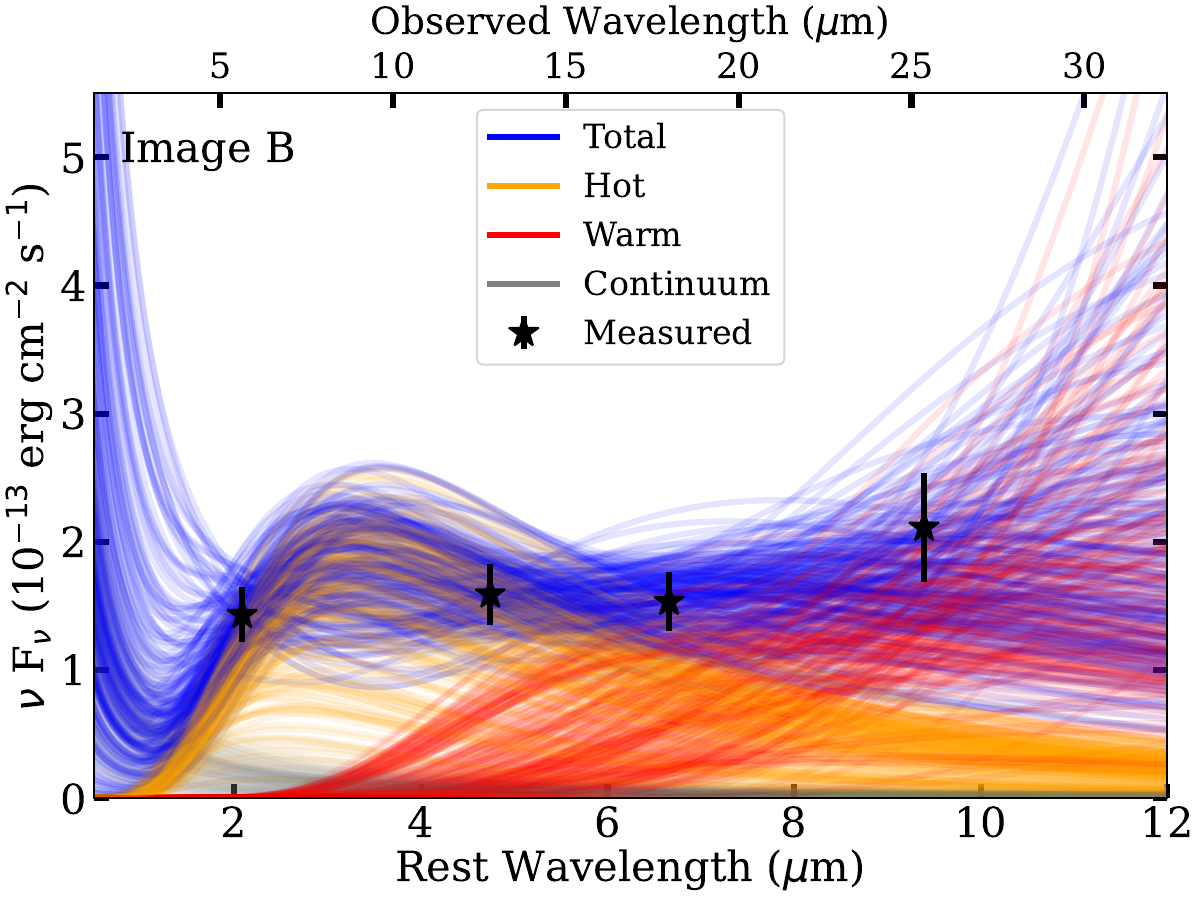}

\hspace{10mm}

\includegraphics[width=0.45\linewidth,trim=0cm 0cm 0cm 0cm,clip]{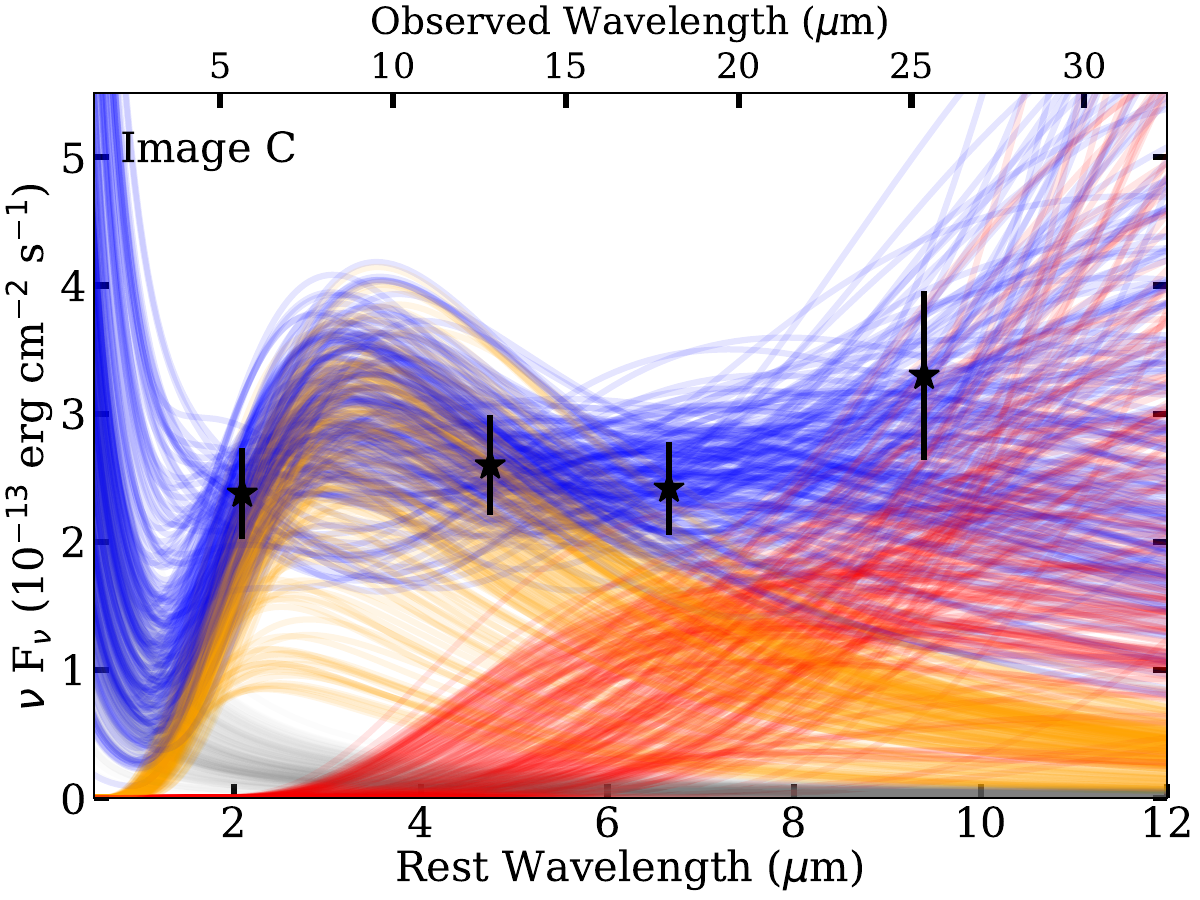}
\vspace{5mm}
\includegraphics[width=0.45\linewidth,trim=0cm 0cm 0cm 0cm,clip]{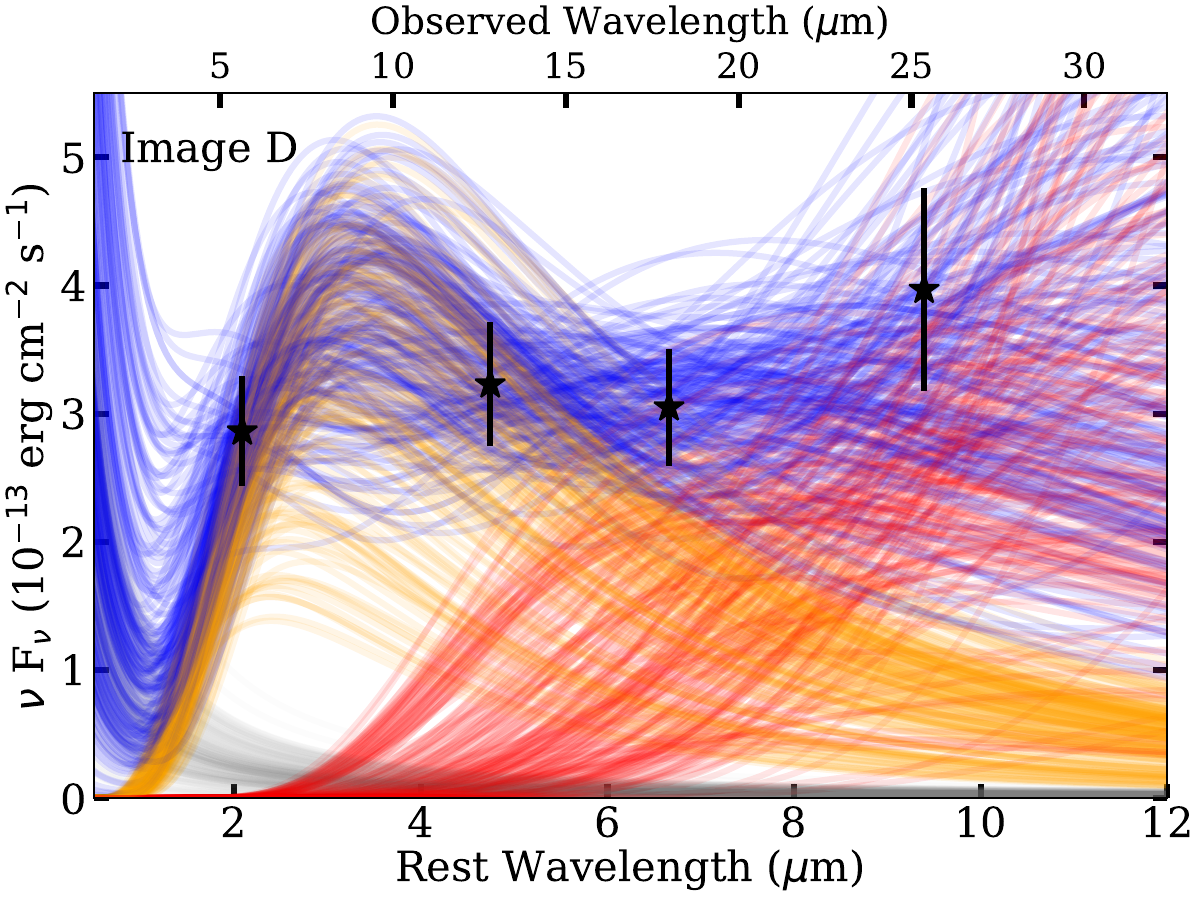}

\caption{Results for SED fitting for separate lensed quasar images fit to a model with continuum plus hot and warm blackbody components with variable temperature. These components represent the quasar accretion disk (gray) and the hot (orange) and warm dust (red) contributions respectively. The amplitude of each model component varied freely between images to accommodate size-dependent microlensing, intrinsic variability, and lensing by the main deflector and potential dark matter substructure. The fits are required to reproduce the observed absolute fluxes as well as the flux ratios in each filter. See Section \ref{sec:sed_fitting} for a description of the model. Each line represents an accepted MCMC draw to illustrate the variations in models.\label{fig:seds} }
\label{fig:prmass}
\end{figure*}

Figure \ref{fig:seds} shows the accepted model drawn from the Markov Chain Monte Carlo, while inferred component flux ratios for the hot and warm dust blackbodies are presented in Table \ref{tab:flux_ratios}. The flux ratios are computed by dividing the normalization of the blackbody component for a given image by the normalization of the corresponding blackbody component for image A. Although we included the continuum emission in our model to estimate the uncertainty it might contribute, we do not present the continuum flux ratios as its contribution to the fluxes was small ($<$10\%) in the observed band-passes.

\begin{table}
\small
\centering
\begin{tabular}{lllllll}
\hline
\hline
  Ratio &  Hot &  Warm       & F2550W     & [OIII] \\
    \hline
B/A &   0.58$^{+0.04}_{-0.07}$      &  0.70$\pm$0.02              & 0.7$\pm 0.007$       &  0.65$\pm 0.04$   \\
C/A &   0.96$^{+0.04}_{-0.07}$     &  1.07$^{+0.03}_{-0.02}$       & 1.05$\pm 0.01$        &  1.25$\pm 0.03$   \\
D/A &   1.23$^{+0.04}_{-0.06}$      &  1.27$\pm$0.02              & 1.27$\pm 0.01$        &  1.17$\pm 0.04$   \\
\hline
\end{tabular}
\caption{Flux ratios and one sigma uncertainties measured through SED fitting, F2550W, and narrow-line [OIII] from  \citet{nierenberg_double_2020}. \label{tab:flux_ratios}}
\end{table}

\section{Discussion}
\label{sec:discussion}

Here we discuss the results of SED fitting and flux ratio measurements in light of other studies of this system.

\subsection{SED Fitting Results}

The hot dust temperature was inferred to be 1200$\pm$100 K, and the warm dust temperature was 300$\pm$100 K. Interestingly, these values are consistent with the best-fit \webbpsf blackbody temperature parameters for F560W (1130 K) and F2550W (250 K, Table \ref{tab:psfPars}). In these filters, the SED model predicts the flux is dominated by the hot and warm dust components respectively. 

The hot dust flux ratios are significantly different from the warm dust flux ratios for images B and C. This is reflected in the flux ratios displayed in Figure 2, which are nearly achromatic for D/A but show small chromatic changes for B and C.  
A microlensing explanation would be consistent with results from \citet{nierenberg_double_2020},
who found clear signatures of microlensing in image C, which had a wider H$\beta$ emission line in C band relative to the other three images. Deformation of the broad emission line profile (such as H$\beta$) is a noted signature of microlensing \citep[e.g.][]{sluse_microlensing_2012,fian_microlensing_2021}, and reflects the differential lensing by stars of the higher velocity wings emitted from the smaller parts of the broad-line region.

From the SED fitting, we see that the flux from the warm dust is dominant relative to the hot dust.
This is consistent with typical quasar SEDs which find a lower covering fraction of hot dust relative to warm dust \citep{mor_hot-dust_2011}. From this result, we expect little contamination in F2550W from the more compact hot dust region. We find that the warm dust flux ratios for this system are consistent with the F2550W flux ratios. 

\subsection{Comparison with past results}

There is a significant difference between the cold dust flux ratios and the [OIII] flux ratios for image C measured by \citet{nierenberg_double_2020}. As discussed in the Introduction, the [OIII] and warm dust emission regions are both extended and not subject to microlensing or time-variability on the day-to-month time scales relevant to galaxy-scale lenses. Therefore, the differences in flux ratios cannot be explained by these phenomena. Furthermore, differential dust extinction is not a likely explanation as the [OIII] emission is redshifted to $\sim$1 $\mu$m at the redshift of the deflector, and the quasar warm torus light is redshifted well beyond this. Assuming all measurement uncertainties have been accurately characterized, we explore two possible explanations below. 

An offset between the centroid of the [OIII] emission and the warm torus emission could create a small difference in the flux ratios. Offsets have been observed to be of order tens of parsecs \citep{singha_close_2022} between the nuclear narrow-line region and the quasar accretion disk. We tested the impact such an offset would make by choosing a macro model that fits the measured image positions and flux ratios, and offsetting the source from the best-fit position. A 10 pc offset, for example, would create a flux-ratio difference of up to 2\% and change the image positions by up to 0\farcs007. However, the flux-ratio changes are not independent of each other and there is no source offset that reproduces both the image positions and flux ratios for the [OIII] and warm dust in this system. Further investigation of the grism data from \cite{nierenberg_double_2020} with simulated offsets between the continuum and the [OIII] region on the two-dimensional grism data would enable limits to be placed on the possible magnitude of such an offset for this system.

Another explanation for the difference in flux ratios is differential milli-lensing by low mass perturbers. The mid-IR and [OIII] sources have intrinsically different characteristic sizes. The two sources could be magnified differently by the same mass perturber. A qualitative example of this effect is provided in Figure~\ref{fig:diff_mag}, in which a small source like the warm torus is strongly de-magnified by a perturbing subhalo, while a larger narrow-line region source is not. As with the example lensed image in Figure \ref{fig:diff_mag}, image C is a saddle image and we would therefore typically expect it to be de-magnified by a local perturbation to the macromodel, thus the observed in the flux ratios could be explained by this type of phenomenon. The differential effect of such a perturbation on the narrow and warm dust flux-ratios would depend on a variety of factors including both the mass of the perturbation and the intrinsic size of the narrow-line region. Based on the grism spectra, \citet{nierenberg_double_2020} placed an approximate upper limit of $\sim$ 100 pc on the FWHM of the narrow-line region for this system based on a lack of differential extension in the spectra of the four lensed images. Such a differential extension would be observed in the grism spectrum if the narrow-line emission was partially resolved \citep[see also][]{nierenberg_probing_2017}. As a test, we started with a macromodel that fits the observed [OIII] flux ratios and image positions. Assuming the [OIII] emitting region has a FWHM of 50 pc, a perturbation with mass scale 10$^7$ M$_\odot$ could reproduce the observed warm torus flux ratios for this system while leaving the [OIII] flux ratios unchanged.
 
In reality, we expect many low-mass halos in the lens and along the line of sight, potentially perturbing all four images simultaneously, therefore we defer a more meaningful physical interpretation of the discrepancy between the [OIII] flux ratios and the mid-IR flux ratios until we have included the effects of full populations of halos and subhalos (Keeley et al. in prep).

\section{Dark matter constraints forecast}
\label{sec:forecast}
Given the flux-ratio precision measured in this work, we can estimate the constraint on dark matter properties obtainable from the full sample based on the scaling simulations by \cite{gilman_probing_2019}. The current WDM constraint is based on a sample of 8 lenses with approximately 6\% measurement precision. Extrapolating to 31 lenses with a 3\% measurement precision for the relative flux ratios yields an estimated 95\% upper limit on a turnover in the half mode mass M$_{\rm{hm}}$ of below 10$^7$ M$_\odot$ if dark matter is cold. This would correspond to a limit on a thermal relic particle mass above 9.7 keV. The current limit from lensing is M$_{\rm{hm}}< 10^{7.8}$ ($M_{\rm WDM}>$5.2 keV) based on 8 lenses with narrow-line measurements \citep{gilman_warm_2020}. Constraining the half mode mass to be below 10$^7$ M$_\odot$ would imply the existence of completely dark subhalos and provide a validation of a major prediction of Cold Dark Matter.

In addition to WDM, \citet{gilman_strong_2021} showed that the compact sources in the JWST dataset make these data highly sensitive to the internal structure of halos. This has particularly relevant consequences for self-interacting dark matter, which can cause halos to undergo core collapse, raising their central densities and therefore their lensing efficiency. Based on the forecasts by \citet{gilman_strong_2021} and the analysis with existing data performed by \citet{gilman_constraining_2023}, the sample size of lenses obtained through this JWST program should enable constraints on self-interaction cross sections in which $> 40 \%$ of halos core collapse. The properties of the SIDM cross section required to produce this quantity of collapsed objects depend on the degree to which tidal stripping and evaporation alter the collapse times for subhalos and on the nature of the self-interaction itself. 
\citet{keeley_pushing_2023} demonstrated that this data set will enable the detection of a mixture of dark matter made of 50\% WDM with half mode mass of 10$^{8.5}$ M$_\odot$ and 50\% CDM.
Similarly, major improvements will be obtained for limits on all dark matter models that produce observed consequences on these scales, including, for example, fuzzy dark matter and PBHs.

\section{Summary}
\label{sec:summary}
We present flux-ratio measurements for DES J0405-3308, the first of 31 systems to be observed in our program to measure rest-frame mid-IR flux ratios of quadruply imaged quasars with JWST.
 
Our main conclusions are as follows:

\begin{enumerate}
    \item We find that the MIRI point spread function is well fit when significant additional jitter is added to the model, and when the source spectrum is treated as a blackbody with variable temperature in each filter.

    \item The flux ratios can be measured to an estimated 6\%, 2\%, 2\%, and 1\%  precision in F560W, F1280W, F1800W, and F2550W, with the dominant source of uncertainty coming from modelling the lensed quasar host galaxy light in the three bluer filters and from the point spread function in F2550W. The absolute flux uncertainties are estimated to be dominated by ongoing instrument calibrations. For this work, we adopt 15\% uncertainties in F560W, F1280W, and F1800W, and 20\% in F2550W, but we expect these to improve in the future. 

    \item We introduce an SED-fitting method that enables us to take into account the high flux-ratio precision and the relatively uncertain absolute flux precision. This model fits for the temperatures of the dust components as well as the relative amplitudes of each component in each lensed image.
    
    \item We estimate the hot and cold dust temperatures for the source to be 1200$\pm$100 K and 300$\pm100$ K. The hot dust region shows substantial microlensing relative to the warm dust region, confirming the sub-parsec size of this region.
    
    \item The flux ratios inferred from the warm dust component of SED fitting are consistent with the flux ratios measured in F2550W. Given current absolute and flux-ratio measurement uncertainties, the warm dust emission flux ratios can be measured to 3\% with one-sigma uncertainty. This sensitivity will enable us to infer population-level statistics of dark matter halos below masses of 10$^7$ M$_\odot$ in future work, thus providing a test of a key prediction of CDM.
    
    \item The F2550W and warm dust flux ratios are inconsistent at a 20\% level with narrow-line flux ratios measured by \citet{nierenberg_double_2020}. This can potentially be explained by the presence of a low-mass dark matter halo magnifying the smaller warm torus light, but not significantly affecting the more extended narrow-line region image fluxes. 
    Full modeling of the substructure and finite size effects, to be presented in a future paper, will be used to study the origin of the discrepancy in more detail.
    
\end{enumerate}

\section*{Acknowledgments}
We thank Crystal Mannfolk, Greg Sloan, and Blair Porterfield for help with observation planning.
We thank Karl Gordon, Mattia Libralato, Jane Morrison, and Sarah Kendrew for their help in answering questions about the data reduction. We thank Marshall Perrin for helpful conversations about {\tt webbPSF}.

This work is based on observations made with the NASA/ESA/CSA James Webb Space Telescope. The data were obtained from the Mikulski Archive for Space Telescopes at the Space Telescope Science Institute, which is operated by the Association of Universities for Research in Astronomy, Inc., under NASA contract NAS 5-03127 for JWST. These observations are associated with program \#2046. 
Support for program \#2046 was provided by NASA through a grant from the Space Telescope Science Institute, which is operated by the Association of Universities for Research in Astronomy, Inc., under NASA contract NAS 5-03127. 

AN and TT acknowledge support from the NSF through AST-2205100 "Collaborative Research: Measuring the physical properties of dark matter with strong gravitational lensing".
The work of LAM and DS was carried out at Jet Propulsion Laboratory, California Institute of Technology, under a contract with NASA. TA acknowledges support the Millennium Science Initiative ICN12\_009 and the ANID BASAL project FB210003. DS acknowledges the support of the Fonds de la Recherche Scientifique-FNRS, Belgium, under grant
No. 4.4503.1. VM acknowledges support from ANID FONDECYT Regular grant number 1231418 and Centro de Astrof\'{\i}sica de Valpara\'{\i}so. 
VNB gratefully acknowledges assistance from a National Science Foundation (NSF) Research at Undergraduate Institutions (RUI) grant AST-1909297. Note that findings and conclusions do not necessarily represent views of the NSF.
KNA is partially supported by the U.S. National Science Foundation (NSF) Theoretical Physics Program, Grants PHY-1915005 and PHY-2210283. AK was supported by the U.S. Department of Energy (DOE) Grant No. DE-SC0009937, by the UC Southern California Hub, with funding from the UC National Laboratories division of the University of California Office
of the President,  by  the World Premier International Research Center Initiative (WPI),  MEXT,  Japan, and by Japan Society for the Promotion of Science (JSPS) KAKENHI Grant No. JP20H05853.
SB acknowledges support from Stony Brook University. DG acknowledges support for this work provided by the Brinson Foundation through a Brinson Prize Fellowship grant, and from the Schmidt Futures organization through a Schmidt AI in Science Fellowship.

\bibliographystyle{apj_2}
\bibliography{references_2}

\label{lastpage}
\bsp
\end{document}